\documentclass[preprint,showpacs,showkeys]{revtex4-2}
\usepackage{graphicx}
\usepackage{amsmath}
\usepackage{amssymb}

\begin{document}

\title{Self-Propelled Collective Motion with Multiplicative Scalar Noise}

\author{Fatemeh Haghsheno and Mohammad Mehrafarin}
\email{mehrafar@aut.ac.ir}
\affiliation{Physics Department, Amirkabir University of Technology, Hafez Ave, Tehran, Iran}

\begin{abstract}

The emergence of order from initial disordered movement in self-propelled collective motion is an instance of nonequilibrium phase transition, which is known to be first order in the thermodynamic limit. Here, we introduce a multiplicative scalar noise model of collective motion as a modification of the original Vicsek model, which  more closely mimics the particles' behavior. We allow for more individual movement in sparsely populated neighborhoods, the mechanism of which is not incorporated in the original Vicsek model. This is especially important in the low velocity and density regime where the probability of a clear neighborhood is relatively high. The modification, thus, removes the shortcoming of the Vicsek model in predicting continuous phase transition in this regime. The onset of collective motion in the proposed model is numerically studied in detail, indicating a first order phase transition in both high and low velocity/density regimes for systems with comparatively smaller size which is computationally desirable.

\end{abstract}
\maketitle

\section{Introduction} 

Nature presents spectacular representations of collectively moving organisms, spanning a wide range of length scales from the size of mammal herds, bird flocks, fish schools and insect swarms, down to size of bacteria colonies, migrating cells and biomolecular micromotors \cite{vicsek2,perc1,perc2}. A common feature characterizing such collective motion and swarming behavior is the macroscopic order they tend to exhibit, in spite of their initial disordered movement. The central question, thus, concerns the nature of this nonequilibrium disorder-order transition and the behavior involved. 

In equilibrium thermodynamics, Mermin and Wagner \cite{mermin} showed that no phase transition into a long range ordered state can occur at nonzero temperature in one or two dimensional systems involving short-range interactions and continuous symmetries. However, far from equilibrium systems can spontaneously break a continuous symmetry (rotational invariance), developing long range order in two dimensions even with only short ranged interactions. Accordingly, the emergence of this contrasting behavior has attracted the attention of physicists during last decades.

Many efforts have been made to determine the general principles governing the appearance of collective order in short range interacting nonequilibrium systems. The earliest attempt seems to be due to Aoki \cite{aoki} who studied the schooling mechanism in fish. Subsequently, Reynolds \cite{reynolds} simulated the motion of some birdlike objects, which he called “boids”, taking into account the following three desiderata:

\noindent (i) Collision avoidance in order to avoid crowding of local flock mates

\noindent (ii) Velocity matching such that objects steer towards the average heading direction of their local flock mates

\noindent (iii) Flock centering in an attempt to stay close to the center of mass of the flock.

\noindent The model was deterministic and had a number of adjustable parameters. Later, to provide a quantitative description of collective motion in presence of perturbations, a particularly simple model was proposed by Vicsek et al. \cite{vicsek}. The Vicsek model (VM) demonstrated a second order phase transition from an initial disordered state to an ordered phase as the level of perturbation decreased.
On the other hand,  Gr\'{e}goire and Chat\'{e} suggested another significant model \cite{gregoire}  by altering the scalar noise term of the VM to a vector noise and obtained a contrary result. They found that in their model, the onset of collective motion is discontinuous meaning that the phase transition is first order.

Following this discrepancy, many attempts were made to substantiate each claim and thus determine the nature of the phase transition in nonequilibrium collective motion \cite{aldana1,aldana2,huepe,baglietto1,aldana3,baglietto2,baglietto3,baglietto4,baglietto5,baglietto6}. In this connection, it was shown that in VM the phase transition  for high velocity and density regime is discontinuous \cite{nagy,vicsek2,chate,chate2,chate3} and that the continuous character in the low velocity/density regime is only an artifact of finite system size.

Here, we introduce a more realistic variant of the VM that is supported by a simple logic,  showing discontinuous phase transition also in the low velocity/density regime for comparatively smaller system sizes. The idea is to consider an uncertainty in the velocity direction of each single neighbor within the interaction circle of the reference particle. This is simply because noise takes into account possible misalignment of the  particle's motion  with the local average velocity. Then, the noise strength will be of the order of square root of the number of particles in the interaction circle, and justifiably vanishes when the vicinity is clear of neighbors so that the particle continues along its current direction. Therefore, it allows for more individual movement in sparsely populated neighborhoods. Thus, we obtain a multiplicative scalar noise that replaces the additive noise of the VM, which is expected to affect the low velocity/density regime behavior where a clear neighborhood is more likely. The consequence is that we now find a first order phase transition for both typical high and low velocity/density regimes.

We give details of our new model with the results of its numerical analysis in Section \ref{new}. First, we briefly review the VM model in the following section.

\section{Additive Scalar Noise Model}\label{models}

In generic analytic models, we have N elements designated as point particles within a two dimensional box of sides $L$ (hence, overall mean density $\rho=N/L^2$), with periodic boundary conditions imposed. All particles move with a fixed speed $v_0$, initially in uncorrelated directions. At each time step $\Delta t$, a reference particle heads  in the average direction of motion of its neighbors lying within a circle of radius $r_0$. The error on the part of the particle to estimate the local average velocity results in slight misalignment, which is taken into account as a random perturbation  by the models. Thus, such models basically differ by their ``noise" content. Below we describe the VM, which was the first swarming model in this context.

Given the position $ x_j(t)$ of the particle $j=1,\dots, N$ and its velocity $\vec{v}_j(t)=v_0\,e ^{i\theta_j(t)}$ at time $t$, the new position after time step $\Delta t$ is calculated from
\begin{equation}\label{eq1}
 \vec{x}_j({t+\Delta{t}})=\vec{x}_j({t})+\vec {v}_j({t})\,\Delta{t}.
\end{equation}
The new velocity direction is determined by updating the angle according to
\begin{equation}
\theta_j({t+\Delta{t}})=\arg \big[\sum\limits_{k \sim j} {e^{i\theta_k({t})}}\big]+\eta \, \xi_j({t})
\label{eq:thetav}
\end{equation}
where the first term yields the average direction of motion of particles neighboring $j$  (including $j$ itself) within  interaction radius $r_0$ at time $t$, and the second term is the additive scalar noise.
$\xi_j$ is a random number chosen from a uniform distribution in the interval $[-\pi , \pi]$, and $0<\eta<1$ is the noise amplitude. Also, in the original VM,  $v_0=0.03$.

As the noise amplitude decreases, or the particle density increases, the system transitions from a disordered state in which the particles move freely in random directions into an ordered state wherein the particles move collectively in the same direction.  The collective behavior of the particles is characterized by the time average, $\langle\varphi\rangle$, of the order parameter
\begin{equation}
\varphi(t)=\frac{1}{Nv_0}| \sum\limits_{j=1}^N {\vec v_j}(t)|
\end{equation}
which corresponds to the magnitude of the normalized average velocity of the particles comprising the system. $\langle\varphi\rangle$  grows to one from zero as $\eta$ is decreased to a critical value $\eta_{\textrm {c}}$, implying that all particles are approximately heading in the same direction.  In VM this phase transition is continuous for low velocities and densities.

The VM can be regarded as the nonequilibrium version of the two dimensional XY model, with the (normalized) velocity as the counterpart of  spin variable and the noise amplitude as temperature. The continuous rotational symmetry and the interaction rule between elements are the same for both models, the only difference being in the dynamics of the VM which updates the interactions at each time step (it corresponds to off-lattice displacement of  elements, which is absent in the XY model). This difference is physically significant and causes distinct behaviors since the XY model does not exhibit a long range ordered phase.

\section{Multiplicative Scalar Noise Model}\label{new}

As mentioned, noise is introduced to take into account possible misalignment of the reference particle's velocity with the average motion in its local vicinity. By considering an uncertainty for the velocity direction of each single neighbor, the noise strength will be $\eta \surd{n_j}$, where $n_j(t)$ is the current number of neighbors of particle $j$ in the vicinity circle (excluding $j$ itself). Then, when the vicinity is devoid of neighbors, there is no cause for error and the particle follows its current course of motion. In other words, we have $\theta_j({t+\Delta{t}})=\theta_j(t)$ when $n_j(t)=0$. This is clearly not the case in Eq.~\eqref{eq:thetav}, where there still remains the additive noise.
 The noise strength is, thus, no longer an externally tunable parameter, but rather depends on the dynamics of the particles. The new multiplicative scalar noise model,  which defines a variant of the VM, is therefore given by the updating rule 
\begin{equation}\label{eq}
\theta_j({t+\Delta{t}})=\arg \big[\sum\limits_{k \sim j} {e^{i\theta_k({t})}}\big]+\eta \, \sqrt{n_j({t})}\, \xi_j({t})
\end{equation}
together with that of  Eq.~\eqref{eq1} for the position variable. 

\subsection{Numerical simulation}

We have studied the behavior of the model defined by Eq.~\eqref{eq} by performing Monte Carlo simulations as a function of two control parameters, namely the particle density  $\rho$, and the noise amplitude $\eta$. Point particles labeled by an integer index j move off the lattice in a two-dimensional box with a velocity of fixed modulus $v_0$.
We applied random initial distribution for the position and velocity of the particles with periodic boundary conditions, having taken $r_0= \Delta{t}=1$ as in the literature.

We thus performed numerical simulations for two different ranges of velocity and density.
To compare our model with the VM we chose these ranges according to \cite{nagy}, where the low velocity regime has been specified by $v_0 \leq 0.1$, and the high velocity regime by $v_0 \geq 0.3$.
In the standard manner, we have considered the variations with $\eta$ of the  time averaged order parameter $\langle\varphi \rangle$ and the Binder cumulant function, $G=1-\langle\varphi^4 \rangle/3{\langle\varphi^2\rangle}^2$, near the transition point.
The binder cumulant is specially useful because it is one of the simplest ratios of moments which takes a universal value at the critical point, where all the curves $G(\eta,L)$ obtained at different system sizes  $L$ cross each other.
A sudden jump in $\langle\varphi \rangle$  near $\eta_c$, accompanied by a minimum in the graph of $G$  (which generally occurs for sufficiently large box size),  signify a first order phase transition.  

\subsubsection{Low velocity and density regime}

\begin{figure}[!h]
\begin{tabular}{cc}
\includegraphics[width=63mm]{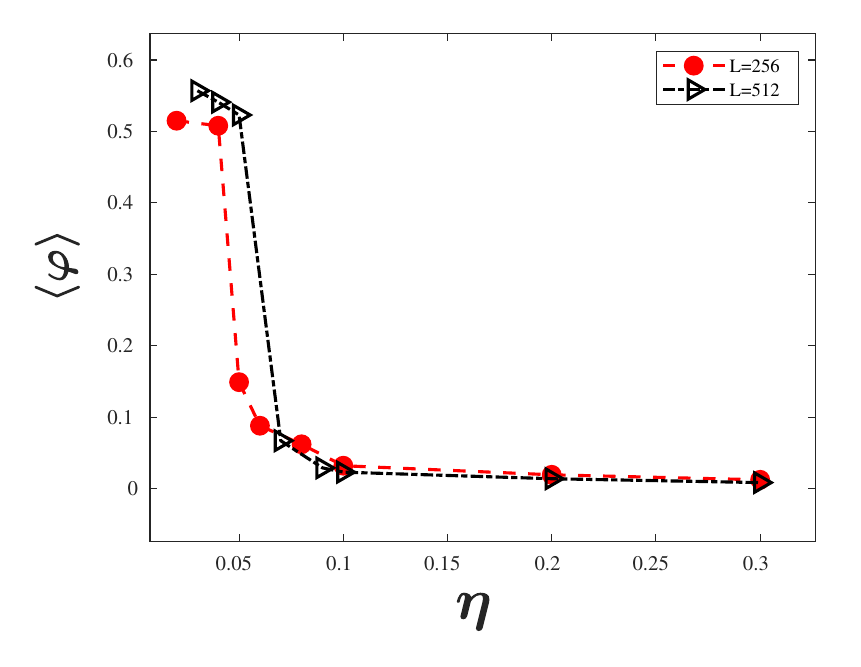}&       
\includegraphics[width=63mm]{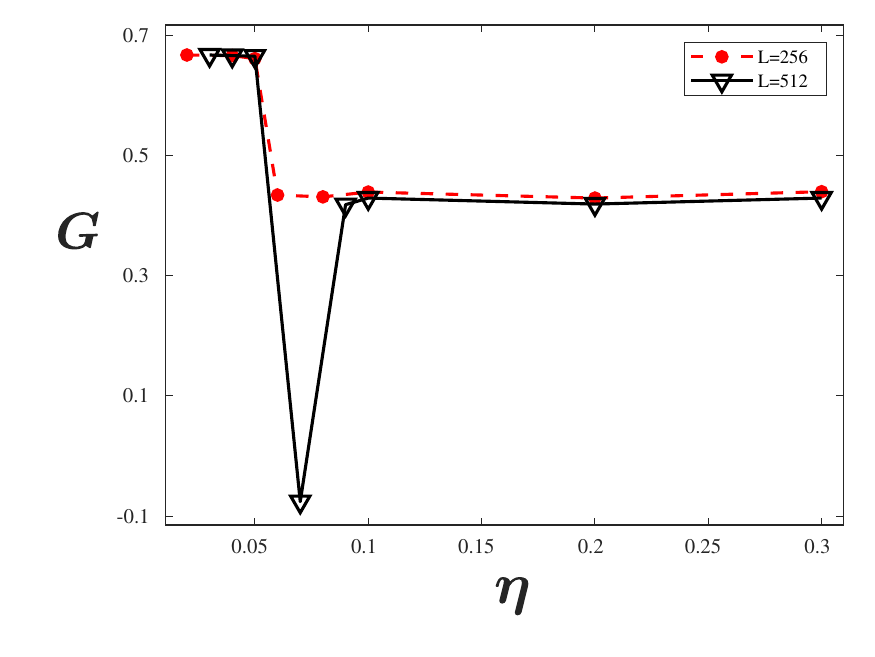}
\end{tabular}
\caption{Discontinuous character of the onset of collective motion for low velocity and density regime specified by  $v_0=0.03, \rho=\frac{1}{8}$, with $L=256, 512$. The time-averaged order parameter $\langle\varphi\rangle$ over $10^6$ time steps (left) and the corresponding Binder cumulant $G$ (right) are plotted as a function of the noise amplitude $\eta$. Discontinuous phase transition is obvious in L=512 in contrast with \cite{nagy} that exhibits a continuous phase transition even in L=512.}
\label{fig1}
\end{figure}

We take $v_0=0.03$ and $\rho=\frac{1}{8}$ corresponding to a small effective step size, $l=v_0\Delta{t}/{r_0}=0.03$. FIG.~\ref{fig1} displays $\langle\varphi \rangle$ as a function of the  noise intensity $\eta$ for a system with $L=256, 512$. There is a discontinuous disorder-order transition in view of the sharp drop of $\langle\varphi \rangle$ at the critical value $\eta_{\textrm{c}}\approx 0.06$. Moreover, the graph of the Binder cumulant $G$, depicted also for the larger size $L=512$ for higher visibility of the minimum, shows the coexistence of two different metastable phases at the transition point. This contrasts with reference \cite{nagy} which reports continuous phase transition in the same regime and system sizes of up to $L=512$.

\subsubsection{High velocity and density regime}

\begin{figure}[!ht]
\begin{tabular}{cc}
\includegraphics[width=63mm]{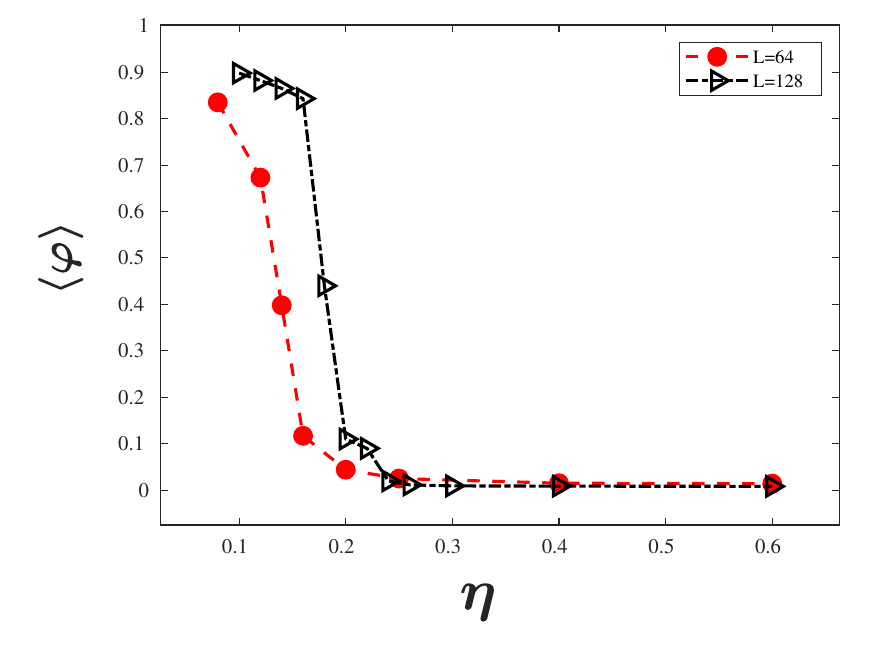}&          
\includegraphics[width=63mm]{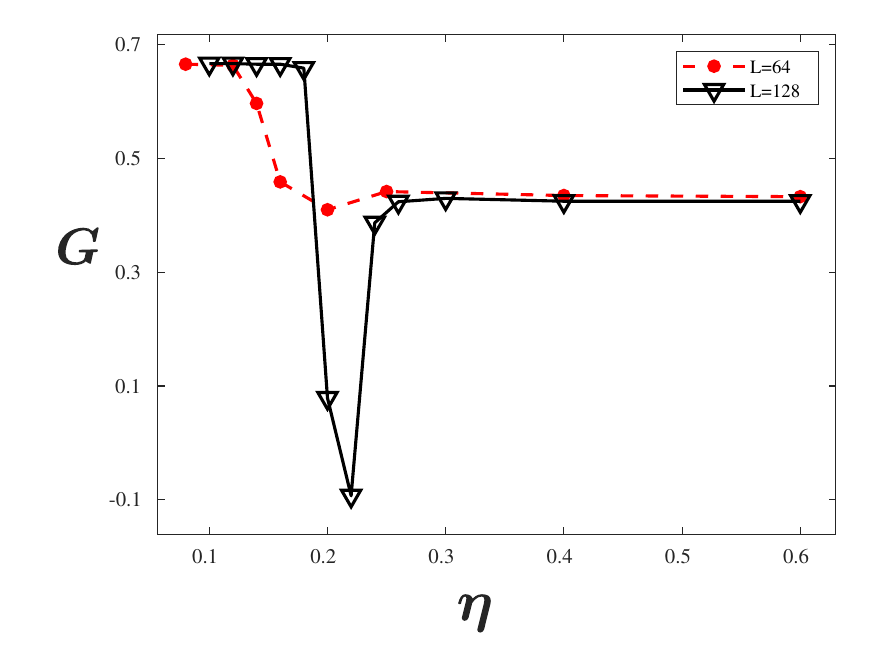}              
\end{tabular}
\includegraphics[width=65mm]{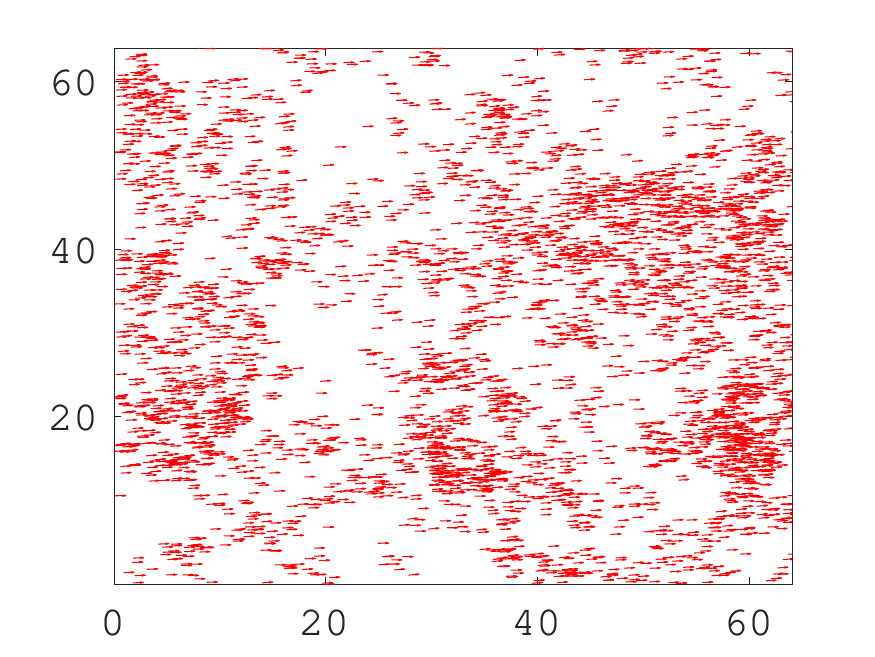}    
\caption {First order transition for high velocity and density regime specified by $v_0=0.5, \rho=1$, with $L=64, 128$. The time-averaged order parameter $\langle\varphi\rangle$ (top left) and the corresponding Binder cumulant $G$ (top right) are plotted as a function of the noise amplitude $\eta$. Also a typical snapshot (bottom) at the transition point is shown with $L=64$ after $10^5$ time steps, where arrows point along the direction of motion. The discontinuous nature of phase transition is clearly observed in L=128, while according to \cite{chate2}, under similar conditions it occurs at larger sizes of the simulation box.}
\label{fig2}
\end{figure}

Simulation results of the multiplicative scalar noise model in high velocity and density regime are presented in FIG.~\ref{fig2}. We started off with $v_0=0.5$ in boxes of size $L=64, 128$ and considered all other conditions as previous, except $\rho=1$. As FIG.~\ref{fig2} (top left) indicates, the phase transition to the state in which all particles are aligned is first order. Another signature of the discontinuous phase transition is given in FIG.~\ref{fig2} (top right) where $G$ exhibits a significant minimum, signaling first order phase transition. The figure also shows in bottom a snapshot of the system with $L=64$ at the transition point  where order sets in. It is appealing that, in our model, the discontinuous phase transition occurs in a smaller system size than that presented in \cite{chate2}.

In conclusion, the  above numerical analysis indicates that our multiplicative noise model yields consistent results for two typical high and low velocity/density regimes considered in the literature, predicting a first order phase transition from disorder to order in self-propelled collective motion. Moreover, the model demonstrates phase transition for comparatively smaller box sizes than required in the VM, particularly in the low velocity/density regime, which facilitates computational work.

\section{Summary}
The emergence of macroscopic order in short range interacting nonequilibrium systems, most notably in self-propelled collective motion, has attracted attention during last decades where many efforts have been made to determine the general principles governing collective order. The first quantitative analysis of collective motion in presence of noise
was presented by Vicsek et al., who predicted continuous transition in the low velocity/density regime. It was later shown that in their model, the phase transition  for high velocity and density regime is discontinuous and the continuous character reported in the low velocity/density regime is only an artifact of finite system size. In this work, we have introduced a more realistic variant of the Vicsek model that is supported by a simple logic,  showing discontinuous phase transition also in the low velocity/density regime. The idea is to allow for more individual movement in lightly populated neiborhoods by considering an uncertainty in the velocity direction of each single neighbor within the interaction circle of the reference particle. The noise strength is of the order of square root of the number of particles in the interaction circle, which therefore vanishes when the vicinity is clear of neighbors so that the particle continues along its current direction. This modification, which yields a multiplicative scalar noise in place of the additive noise of the Vicsek model, is especially important in the low velocity/density regime where a clear neighborhood is more likely. Our numerical analysis yields consistent results in two typical high and low velocity/density regimes indicating a first order phase transition with comparatively smaller system size.

\end{document}